\documentclass[aps,pre,twocolumn,float]{revtex4}
\usepackage{amsmath,bm,epsfig}

\def\Fbox#1{\vskip1ex\hbox to 8.5cm{\hfil\fboxsep0.3cm\fbox{%
  \parbox{8.0cm}{#1}}\hfil}\vskip1ex\noindent}  

\newcommand{\sFrac}[2]{{\textstyle\frac{#1}{#2}}}
\let \= \equiv \let\*\cdot \let\~\widetilde \let\-\overline

\begin{document}
\title{An Athermal Brittle to Ductile Transition in Amorphous Solids}
\author{Olivier Dauchot$^{1,2}$, Smarajit Karmakar$^1$, Itamar Procaccia$^1$ and Jacques Zylberg$^1$}
\affiliation{$^1$Department of Chemical Physics, The Weizmann
 Institute of Science, Rehovot 76100, Israel \\$^2$CEA - Saclay
F-91191 Gif sur Yvette Cedex France.}
\date{\today}
\begin{abstract}
Brittle materials exhibit sharp dynamical fractures when meeting Griffith's criterion, whereas
ductile materials blunt a sharp crack by plastic responses. Upon continuous pulling ductile materials exhibit a necking instability which is dominated by a plastic flow. Usually one discusses the brittle to ductile transition as a function of increasing temperature. We
introduce an athermal brittle to ductile transition as a function of the cut-off length of the inter-particle potential. On the basis of extensive numerical simulations of the response to pulling the material boundaries at a constant speed we offer an explanation of the onset of ductility via the increase
in the density of plastic modes as a function of the potential cutoff length. Finally we can resolve an old riddle: in experiments brittle
materials can be strained under grip boundary conditions, and exhibit a dynamic crack when cut
with a sufficiently long initial slot. Mysteriously, in molecular dynamics simulations it appeared that
cracks refused to propagate dynamically under grip boundary conditions, and continuous pulling
was necessary to achieve fracture. We argue that this mystery is removed when one understands the
distinction between brittle and ductile athermal amorphous materials.
\end{abstract}
\maketitle

\begin{figure}
\includegraphics[scale = 0.5]{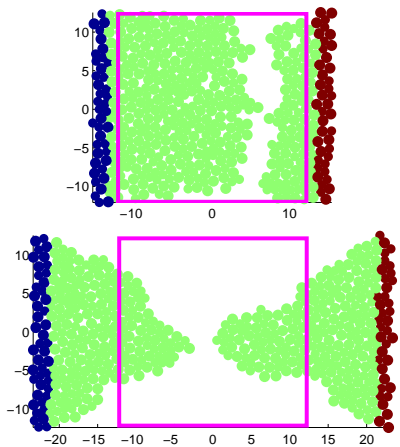}
\caption{Color online: The result of pulling quasi-statically the side boundaries (the particles in the side walls are
held frozen). Upper panel: the material is ``brittle" and the pulling results in a clean crack. Lower panel:
the material is ``ductile" and the pulling induces a plastic flow with a necking instability. Note the large difference in the relative increase in system length compared to the original system size in the two panels. The number of
particles $N=800$}
\label{britvsduct}
\end{figure}

\section{Introduction}
Consider the following numerical experiment, which is
also exhibited in Fig. \ref{britvsduct} and in the movies available in \cite{movies}.
A piece of amorphous solid at rest is contained in the square (pink)
box as plotted. The material is prepared by quenching a
binary liquid of point particles (see potentials below), and
removing the periodic boundaries. The resulting system
has side walls consisting of pinned particles (identical to the bulk particles, blue and brown
in Fig. \ref{britvsduct}) forming a rigid body which is pulled,
the upper and lower boundaries are free. The walls are
moved by an increment equal in size and opposite in sign.

In the upper panel the material is brittle, and cannot deform much before it fractures. In the lower panel the
material is ductile, and the distance between the walls
increases by about a factor of two before the plastic deformation that is induced creates a thin neck that is eventually broken. The huge difference in response is obtained
with materials in the two panels that are identical in every respect except for the cutoff length of the attractive
part of their inter-particle potentials, see Fig. \ref{potentials}. The aim
of this paper is to explore the intrinsic relation between
the microscopic properties of the inter-particle potential
and the brittle to ductile transition. We explicitly do not
consider in this paper the interesting effects of temperature, aspect ratio, system size and dimensionality, which
of course are also important in determining a material's
brittleness.

One of our main motivations in this paper is a mystery
that had clouded the fracture literature for a number of
years now. The issue is crack propagation in amorphous
solids under grip boundary conditions. In experiments
one can pull the boundaries of a given sample and grip
them, and then cut a slot at the edge of the sample \cite{98HM}. In
brittle materials this leads to a fast development of the
a dynamic fracture (when the length of the initial slot
exceeds the Griffith's length \cite{98Fre}). For some unknown reason
one always failed to repeat such experiments in direct molecular dynamic simulations, leading recently to
some claims that only by defining `bonds' between particles and letting these bonds `break' under some criteria
one can sustain a dynamical brittle fracture \cite{11HKL}. Obviously
this is not a satisfactory answer; here we argue that the
mystery is removed when one understands what makes a
material brittle or ductile, and in truly brittle materials
there is no problem in seeing a dynamical fracture under
grip boundary conditions.

In this paper we want to understand how the response of an amorphous system subjected
to uniaxial load depends on the parameters of the potential, and in particular the cut-off length $r_{\rm co}$.
This will provide us with a fair insight into the athermal
brittle to ductile transition. We will relate the degree of ductility to the
growing importance of the contribution of "plastic modes" to the density of states of the material.

The structure of this paper is as follows: In Sect. \ref{systems} we describe the glassy system used and how we control the
range of the inter-particle potential. The numerical experiments and their results are described in Sect. \ref{experiments}.
In Sect. \ref{plastic} we relate and explain the findings of Sect. \ref{experiments} to the density of states of the glassy
materials, explaining that with increasing the range of the potential one changes the weight of the contribution of the plastic
modes in the density of states. This correlates very well with the tendency towards ductility as seen in the numerical experiments. Finally
in Sect.\ref{fracture} we focus on the riddle of brittle fractures and show that there is no riddle at all - one simply needs to
make the material brittle enough to allow brittle fractures to run spontaneously under grip boundary conditions. In Sect. \ref{summary} we
conclude the paper and reiterate the dramatic effect of the inter-particle's potential properties on the behavior of a material under
load \cite{11KLPZ}.

\section{System and Methods}
\label{systems}
\begin{figure}
\centering
\hskip -1. cm
\includegraphics[scale = 0.50]{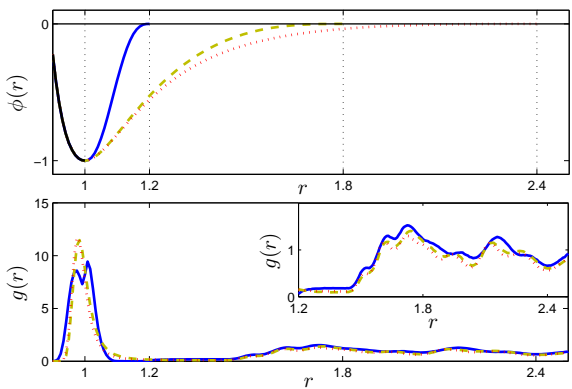}
\caption{Color online: Upper panel: Examples of three different potentials with $r_{\rm co}= 1.2, 1.8$ and 2.4 used in our numerical experiments. The first and the third correspond to the experiment described in Fig.~\ref{britvsduct}. Lower panel: the pair radial
distribution function $g(r)$ for the three corresponding potentials in the upper panel. Note that the second shell of
neighbors begins around $r=1.5$}
\label{potentials}
\end{figure}

For the numerical experiments we employ a generic glass former in 2-dimensions in the form of a binary mixture whose amount of bi-dispersity
of `small' and `large' particles was chosen to avoid any crystallization. In fact the particles interact
by inter-particle potentials as shown in Fig. \ref{potentials}, with the analytic form
\begin{widetext}
\begin{equation}\label{potential}
\phi\!\left(\!\sFrac{r_{ij}}{\lambda_{ij}}\!\right) =
\left\{\begin{array}{ccl}
&&4\varepsilon\left[\left(\frac{\lambda_{ij}}{ r_{ij}}\right)^{12}-\left(\frac{\lambda_{ij}}{r_{ij}}\right)^{6}\right]
\ , \hskip 3.5 cm\frac{r_{ij}}{\lambda_{ij}}\le \frac{r_{\rm min}}{\lambda}\\
&&\varepsilon\left[a\left(\frac{\lambda_{ij}}{ r_{ij}}\right)^{12}-b\left(\frac{\lambda_{ij}}{ r_{ij}}\right)^{6}
+ \displaystyle{\sum_{\ell=0}^{3}}c_{2\ell}\left(\sFrac{r_{ij}}{\lambda_{ij}}\right)^{2\ell}\right]
  \ , \hskip 0.5 cm\frac{r_{\rm min}}{\lambda}<\frac{r_{ij}}{\lambda_{ij}}<\frac{r_{\rm co}}{\lambda} \\
&&\quad\quad\quad\quad 0
   \ , \hskip 5.4 cm  \frac{r_{ij}}{\lambda_{ij}} \ge\frac{r_{\rm co}}{\lambda}
\end{array}
\right.\!\!
\end{equation}
\end{widetext}
Here $r_{\rm min}/\lambda_{ij}$ is the length where the potential attain it's minimum, and $r_{\rm co}/\lambda_{ij}$
is the cut-off length for which the potential vanishes. The coefficients $a,~b$ and $c_{2\ell}$ are chosen such
that the repulsive and attractive parts of the potential are continuous with two derivatives at the potential minimum and the potential goes
to zero continuously at $r_{\rm co}/\lambda_{ij}$ with two continuous derivatives as well.
The interaction length-scale $\lambda_{ij}$ between any two particles $i$ and $j$ is $\lambda_{ij} = 1.0\lambda$, $\lambda_{ij} = 1.18\lambda$
and $\lambda_{ij} = 1.4\lambda$ for two `small' particles, one `large' and one `small' particle and two `large' particle respectively. The
unit of length $\lambda$ is set to be the interaction length scale of two small particles, $\varepsilon$ is the unit of energy and $k_B = 1$.

The systems employed in the pulling experiments consist of configurations which were quenched from systems initially equilibrated at high temperature and pressure ($T=1.0,\ P=1.0$) with periodic boundary conditions. The number of particles $N$ varied between $N=800$ and $N=67000$. We used a modified Berendsen thermostat which couples a constant number of particles to the bath, regardless of the system size \cite{10KLPZ}. The systems are cooled to temperature $T=10^{-2}$ while keeping the pressure high, in order to avoid the creation of any
holes in the material. Then the pressure is reduced to zero, $P=0.0$, such that the periodic boundary conditions could be removed;  particles forming the right and left walls were frozen, but the upper and lower boundaries were rendered free.

From this point the system was treated differently for the brittle crack and the necking experiments. The brittle
crack experiment starts by loading the system uniaxially (with a constant velocity such that $v_{\rm wall} \ll c_s/10$) until a desired stress
is reached, and then the side walls are held fixed. A cut is then implemented by the cancelation of forces
crossing an imaginary line of desired length which starts at the lower boundary. The evolution of the crack is simulated by molecular dynamics at $T = 0.0$ and requires no further loading of
the system.

The necking experiments were conducted in athermal
quasi-static conditions. For those experiments the systems were further minimized with respect to the acting
forces on each particle, making them athermal. The right
and left walls were then moved with a strain increment
$\delta \ell = 5 \cdot 10^{-3}$. The response we investigate is the relative increase in the system's length $\Gamma = \Delta L/L$ along the
stretching direction. The maximal value that the system
is able to accommodate before it breaks into two parts is
denoted as $\Gamma_{\rm max}$ \cite{footnote}.

\section{Numerical Experiments and Results}
\label{experiments}
\begin{figure}
\centering
\hskip -1. cm
\includegraphics[scale = 0.55]{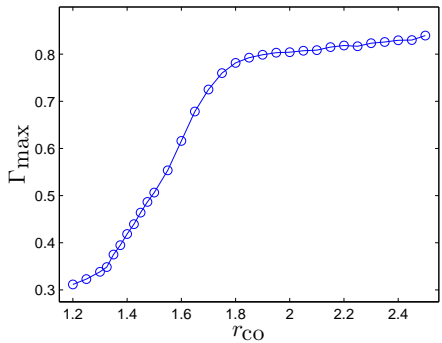}
\caption{Color online: The dependence of $\Gamma_{\rm max}$ on $r_{\rm co}$ for the geometry of Fig. \ref{britvsduct} (square
initial conditions). Note that $\Gamma_{\rm max}$ tends to saturate at a given value $\Gamma^\infty_{\rm max}$ which becomes independent on $r_{\rm co}$; in athermal conditions it depends on the geometry of the sample only.}
\label{bare}
\end{figure}

In this paper we describe only 2-dimensional simulations with a square initial geometry. In fact  we have set up samples of height $H$ and length $L=aH$, with
$a\in [1,5]$, but we did not find any exciting aspect-ratio dependence beyond the obvious. For the square initial geometry the athermal response was characterized by $\Gamma_{\rm max}$. In Fig. \ref{bare} we exhibit the dependence of
$\Gamma_{\rm max}$ on $r_{\rm co}$ for square systems $(a=1)$.
It is sufficient to limit the range of the potential cut-off to
\begin{equation}
r_{\rm co} \in  [1.2\lambda;\ 2.5\lambda]. \label{lengths}
\end{equation}
The lower limit for $r_{\rm co}$ is dictated by the position of the first peak in $g(r)$ cf. Fig. \ref{potentials} lower panel. For $r_{\rm co} <1.2$ the potential gets shorter than the first shell of neighbors. The upper limit is determined by the convergence of our 1-parameter family of potentials to the asymptotic
Lennard-Jones potential.
Within this range of potentials $\langle \Gamma_{\rm max} \rangle$ was evaluated by averaging over 4500 samples. It is found to  increase smoothly  with $r_{\rm co}$ reaching a limiting value $\Gamma^\infty_{\rm max}$.  When the system reaches the limiting value
$\Gamma^\infty_{\rm max}$ we refer to it as ``maximally ductile".

The qualitative understanding of the result shown in Fig.~\ref{bare} is quite straightforward. In the limit $r_{\rm co}\to
r_{\rm min}$ any cavity that forms whose dimensions is larger than $r_{\rm co}$ will not heal since there are no
attractive forces operating across it. Rather, it will propagate rapidly to span the system in a form of a brittle fracture.
With $r_{\rm co}$ increasing, a small cavity will still be held by attractive forces across it. The higher is $r_{\rm co}$
the larger is the number of neighbors that each particles feels attracted to, and one needs to create a very large
cavity indeed to overcome these attractive forces. On the other hand when $r_{\rm co}$ increases beyond the second shell
of the binary correlation function, not much can change. With our choice of potentials, in the limit $r_{\rm co}\to \infty$
we simply converge to the infinite-range Lennard-Jones potential which has a negligible attractive force beyond, say, $r/\lambda=3$.
This qualitative understanding is further supported by examining the contribution of the plastic modes to the density of
states, as we do next.

\section{The contribution of plastic modes}
\label{plastic}

It is well known that in amorphous solids the density of states deviates from the classical Debye formula which was developed for
a perfect elastic solid. The deviations were referred to as the 'Boson peak' but very little reliable analytic statements were
made about the form of this deviation \cite{09IPRS}. In recent work of the present group it was proposed that one can provide a useful analytic
form of the density of states (the HKLP model \cite{11HKLP}) which will be particularly useful at the low end tail of frequencies. In terms of the eigenvalues of the Hessian matrix $\lambda_i=\omega^2_i$ where $\omega_i$ is the $i$'th normal mode frequency the proposed model is
\begin{equation}
P\left(\frac{\lambda}{\lambda_D}\right) =Nd \left[ \frac{d}{2} \left(\frac{\lambda}{\lambda_D}\right)^{(d-2)/2} + B \left(\frac{\lambda}{\lambda_D}\right)^\theta \right] \ , \label{HKLPmodel}
\end{equation}
where the first term on the RHS is the Debye formula in which $N$ is the number of particles, $d$ the space dimension and
$\lambda_D$ is the Debye cutoff frequency which in 2-dimensions reads $\lambda_D= 8\pi \mu$ with $\mu$ being the shear modulus.
The second term is our proposed addition to the model density of states that should be appropriate for all generic amorphous
solids in the limit $\lambda\ll \lambda_D$. This term contains contributions from eigenvalues $\lambda_i$ that depend on the
external mechanical loading, and go to zero with increasing the load, harking the appearance of plastic instabilities \cite{11HKLP}. The exponent $\theta$ in this plastic contribution is not universal, it depends on the amount of disorder in the glassy system, and it was shown to be an important characteristic of the mechanical properties of amorphous solids \cite{11HKLP}.

In numerical calculations it is easier to evaluate the cumulative distribution $C(\lambda)\equiv \int_0^\lambda P(\tilde \lambda) d\tilde \lambda$, whose form is
\begin{equation}
C\left(\frac{\lambda}{\lambda_D}\right) = Nd\left[ \left(\frac{\lambda}{\lambda_D}\right)^{d/2} + \frac{B}{1+\theta} \left(\frac{\lambda}{\lambda_D}\right)^{1+\theta}      \right] \ . \label{cum}
\end{equation}

\begin{figure}
\centering
\hskip -1. cm
\includegraphics[scale = 0.58]{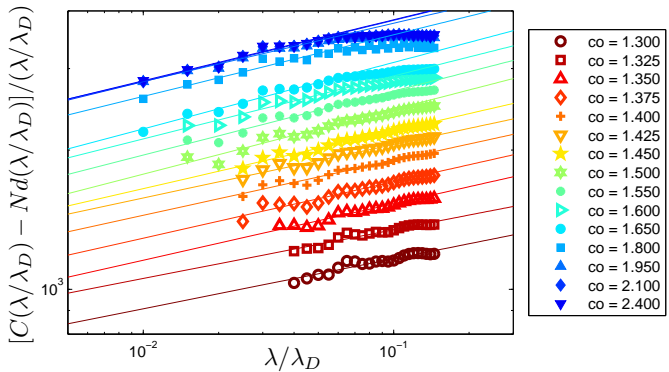}
\caption{Color online: the cumulative density of states from which the Debye contribution was subtracted. The remaining plastic contribution
is plotted in log-log coordinates such that the intercept is $B/(1+\theta)$ and the slope is $\theta$.  }
\label{Clam}
\end{figure}

\begin{figure}
\centering
\hskip -1. cm
\includegraphics[scale = 0.6]{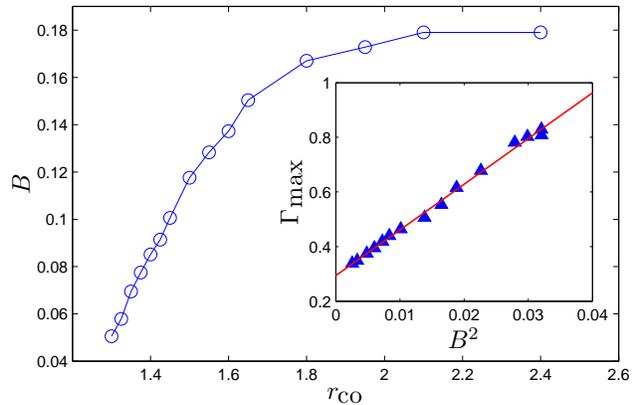}
\caption{Color online: the value of the weight $B$ in Eq. \ref{cum} as a function of $r_{\rm co}$. Inset: $\Gamma_{\rm max}$ vs. $B^2$.}
\label{correl}
\end{figure}

In the present context we wish to understand the relation between the tendency towards ductility as $r_{\rm co}$ increases to the
amount of plasticity as expressed, say, by the value of the weight coefficient $B$ and the exponent $\theta$. To this aim
we have computed the cumulative density of states $C\left(\frac{\lambda}{\lambda_D}\right)$ and subtracted from it the exactly
known Debye contribution. The resulting plots are shown in Fig. \ref{Clam}. We see that the model \ref{HKLPmodel} is faithfully
representing the data for small values of $r_{\rm co}$, whereas for the largest ones the amount of plastic modes increases beyond the first
term contribution in the model. This is however not causing any difficulty in estimating the weight $B$ (the intercept in Fig. \ref{Clam}) which is shown in Fig. \ref{correl}.

It is quite obvious that our $\Gamma_{\rm max}$ dependence on $r_{\rm co}$ (cf. Fig. \ref{bare}) is qualitatively similar to the dependence of the weight $B$ on $r_{\rm co}$. It is therefore natural to plot $\Gamma_{\rm max}$ as a function of $B$. A convincing linear dependence can be obtained by plotting $\Gamma_{\rm max}$ as a function of $B^2$ as is shown in Fig. \ref{correl}.

\section{The brittle crack}
\label{fracture}

\begin{figure}
\centering
\hskip -1. cm
\includegraphics[scale = 1.0]{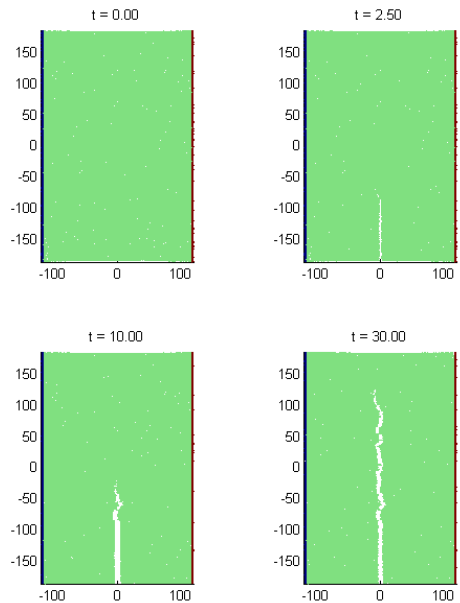}
\caption{Color online: the development of a brittle crack under grip boundary conditions. The finite size effect appears as
bending of the upper boundary, releasing the stress before the fracture hits the boundary.}
\label{brittle}
\end{figure}
\begin{figure}
\centering
\hskip -1. cm
\includegraphics[scale = 0.5]{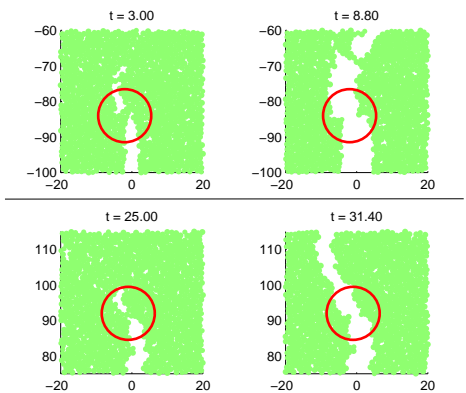}
\caption{Color online: the development of a brittle crack appears many times to consist of the creation of a cavity {\em ahead of the crack},
followed by a necking process in which the cavity and the body of the crack coalesce. Shown are two such events, where the
left panels exhibit the cavities ahead of the crack, and the right panels exhibit the crack after the necking events coalesced
the crack with the cavities.}
\label{cavities}
\end{figure}

We are now able to construct a system where all the microscopic parameters are tuned to minimize the occurrence
of plastic events, a system which can be denoted `maximally brittle'. We expect such a system to fail under grip boundary conditions upon the
application of a cut in one of the free boundaries. For
a large value of $r_{\rm co}$, a small cavity can still be held by
attractive forces across it. The longer is $r_{\rm co}$ the larger
is the number of neighbors that each particles feels attracted to, and one needs to create a very large cavity
indeed to overcome these attractive forces. On the other
hand, in the limit of $r_{\rm co}\to r_{\rm min}$ any cavity that forms
whose dimensions is larger than $r_{\rm co}$ cannot heal since
there are no attractive forces operating across it. Rather,
it would propagate rapidly to span the system in a form of
a brittle fracture. The system we used
to demonstrate this concept is composed of $N = 67000$
particles with a cut-off length of $r_{\rm co}=1.20$ and $\varepsilon = 10.0$; this choice was made to increase the
stability of the system and prevents any unwanted events during the loading part of the simulation.
The experiment is run by means of molecular dynamics at $T = 0.0$. The
size of the cut is 100$\lambda$. The evolution of the crack can be
appreciated in Fig. \ref{brittle}.

When the fracture originates from a cut,
the system does not need to create a cavity in order to propagate a crack. The stress field around the tip of the cut is greatly enhanced by the strong curvature of the boundary, and once the Griffith's criterion is fulfilled the crack can run. Nevertheless during the propagation of the crack we observe many times the formation of a cavity ahead of the crack tip followed
by necking to coalesce the cavity with the body of the crack. The observed dynamics is presented in Fig. \ref{cavities}, which employs
snapshots from the brittle crack development that is shown in Fig. \ref{brittle}. We note in passing that this mechanism
of propagation of brittle cracks was seen experimentally in Ref. \cite{99BP}. In Ref. \cite{04BMP} this same mechanism was proposed
as the reason for the roughening of the fracture surface. This roughening is seen to the bare eye in Fig. \ref{brittle}; its
quantitative assessment and its relation to the mechanism in question will be considered in a future publication.

\section{summary and conclusions}
\label{summary}

The main result of this paper is the dramatic change in material mechanical properties that results from changing
the inter-particle potential interaction length. We introduced the athermal brittle to ductile transition, showing
that even at $T=0$ we can go all the way from brittle to ductile response only by changing the cutoff length $r_{\rm co}$.
There is an interesting analogy between increasing the cutoff length and raising the temperature; in both case the system
becomes `softer'. This softening can be quantitatively described by our model of density of states of amorphous solids,
in which the coefficient $B$ measures the increasing weight of plastic modes, which are those eigenfunctions of the Hessian
matrix whose eigenvalues vanish in saddle-node bifurcations when the system is strained \cite{11HKLP}. Indeed, we found
a linear dependence between the degree of lengthening of the system before breaking and $B^2$. Finally, the new understanding of brittleness
vs. ductility allowed us to disperse a long standing conundrum - how to get a brittle crack running under grip boundary
conditions. It turns out that previous attempts failed simply because the system chosen were not brittle enough. Building a
`maximally brittle' material results in a nice running crack under grip boundary conditions as seen in many experiments.

\acknowledgments
This work had been supported in part by an advanced ``ideas" grant of the European Research Council, the Israel Science
Foundation and the German Israeli Foundation. We thank Edan Lerner for useful discussions.


\begin{thebibliography}{99}
\bibitem{footnote}
There exist materials in which the necking instability results in a very thin thread that connects the tips
of the triangular shapes which are left behind. We do not observe such threads in our simulations and our value
of $\gamma_{\rm max}$ should be understood without the formation of atomic chains.
\end{thebibliography}
\end{document}